\documentclass[twocolumn,aps,superscriptaddress,showpacs]{revtex4}

\usepackage{amssymb}
\usepackage{amsmath}
\usepackage{graphicx}
\usepackage[normalem]{ulem}
\usepackage[dvips]{color}

\begin{document}

\title{Investigating different $\Lambda$ and $\bar\Lambda$ polarizations in relativistic heavy-ion collisions}
\author{Zhang-Zhu Han}
\affiliation{Shanghai Institute of Applied Physics, Chinese Academy
of Sciences, Shanghai 201800, China}
\affiliation{University of Chinese Academy of Sciences, Beijing 100049, China}
\author{Jun Xu\footnote{corresponding author: xujun@sinap.ac.cn}}
\affiliation{Shanghai Advanced Research Institute, Chinese Academy of Sciences, Shanghai 201210, China}
\affiliation{Shanghai Institute of Applied Physics, Chinese Academy
of Sciences, Shanghai 201800, China}

\date{\today}

\begin{abstract}
Based on the chiral kinetic equations of motion, spin polarizations of various quarks, due to the magnetic field induced by spectator protons as well as the quark-antiquark vector interaction, are studied within a partonic transport approach. Although the magnetic field in QGP enhances the splitting of the spin polarizations of partons compared to the results under the magnetic field in vacuum, the spin polarizations of $s$ and $\bar s$ quarks are also sensitive to the quark-antiquark vector interaction, challenging that the different $\Lambda$ and $\bar \Lambda$ spin polarization is a good measure of the magnetic field in relativistic heavy-ion collisions. It is also found that there is no way to obtain the large splitting of the spin polarization between $\Lambda$ and $\bar \Lambda$ at $\sqrt{s_{NN}}=7.7$ GeV with partonic dynamics.
\end{abstract}


\maketitle

Understanding the properties of the quark-gluon plasma (QGP) is one
of the main purposes of relativistic heavy-ion collision
experiments. In noncentral heavy-ion collisions, QGP is expected to
be polarized perpendicular to the reaction
plane~\cite{Gao08,Huang11,Jia16} due to the large angular momentum
as well as the strong magnetic field. Theoretical studies predict
that the strong vorticity and magnetic field lead to a series of
chiral effects (see, e.g., Ref.~\cite{Khar16} for a review) as well as the
spin polarizations of hyperons and vector
mesons~\cite{Liang05,Lia05,Bar07,Beca13}, which are experimentally
measurable through their decays. On the experimental side,
continuous efforts have been made on measuring the spin polarization
of these particles~\cite{STAR07,STAR08,STAR18,Lisa16}. In the collision systems at
higher energies with nearly zero baryon chemical potential, shorter
duration of the magnetic field, and smaller angular velocitiy, the
spin polarizations of $\Lambda$ and $\bar{\Lambda}$ are found to be
very small~\cite{STAR07,STAR18}. Recently, the finite spin
polarizations of $\Lambda$ and $\bar{\Lambda}$ at lower collision
energies have been observed experimentally~\cite{Lisa16}, with the
$\bar{\Lambda}$ spin polarization slightly larger than that of
$\Lambda$. Considerable efforts have been devoted to understanding
the polarization of
$\Lambda$~\cite{Ari16,Pang16,Kar17,LiHui17,Sun17} but few of them
try to address the different spin polarizations of $\Lambda$ and $\bar{\Lambda}$.

The studies in Refs.~\cite{Ari16,Pang16,Kar17,LiHui17,Sun17} attribute the hyperon
polarization to the coupling to the vorticity field of the QGP, and the spin polarizations of quarks and antiquarks are affected in a similar way. On the other hand, the vector potentials, including those from
the quark-antiquark vector interaction and the electromagnetic
field, are expected to be responsible for the different
polarizations for $\Lambda$ and $\bar\Lambda$ at lower collision
energies. Due to the finite baryon chemical potential, quarks and
antiquarks are affected by different spin-dependent interactions in the
baryon-rich matter. It was also proposed that the difference of the spin polarization between $\Lambda$ and $\bar\Lambda$ can be used as a measure of the magnetic field in relativistic heavy-ion collisions (see, e.g., Ref.~\cite{Mul18}), with the strength of the later suffering from the uncertainty of the electrical conductivity of the QGP. The strength of the vector potentials, especially the magnetic field, is responsible for the occurrence of the chiral magnetic effect and the chiral magnetic wave.

In the present study, we investigate the different spin polarizations of $\Lambda$ and $\bar\Lambda$ in Au+Au collisions at $\sqrt{s_{NN}}=39$ and 7.7 GeV as a result of the
vector potentials with partonic transport simulations based on the chiral kinetic equations of motion. The vector potentials include the dominating magnetic field contribution from the spectator protons in the QGP with a temperature-dependent electrical conductivity, and the space component of the quark-antiquark vector potential related to the net quark flux. We found that the $s$ and $\bar{s}$ quark spin polarizations, which are responsible for the $\Lambda$ and $\bar\Lambda$ spin polarizations via the coalescence model, are sensitive to the strength of both the magnetic field and the quark-antiquark vector potential. In addition, there is no way to generate a large splitting of the spin polarization between $\Lambda$ and $\bar\Lambda$ at $\sqrt{s_{NN}}=7.7$ GeV with partonic dynamics.

We generate the initial phase-space information of partons from a multiphase transport (AMPT) model~\cite{Lin05}, with the momenta of initial partons from melting hadrons produced by the Heavy-Ion Jet INteraction Generator (HIJING) model~\cite{hijing}, and their coordinates in the transverse plane $(x,y)$ set to be the same as those of the colliding nucleons that produce their parent hadrons. In order to take into account the finite thickness of the QGP medium at $\sqrt{s_{NN}}=39$ and 7.7 GeV, the longitudinal coordinates ($z$) of initial partons are sampled uniformly within $(-lm_N/\sqrt{s_{NN}}, lm_N/\sqrt{s_{NN}})$, where $l=14$ fm is approximately the diameter of the Au nucleus, and $m_N=0.938$ GeV is the nucleon mass. Each parton is given a formation time related to the energy and the transverse mass of its parent hadron~\cite{Lin05}. Afterwards, the evolution of these partons is described by transport simulations with elastic scatterings between all partons, with the isotropic cross section of 3 mb at 7.7 GeV and 10 mb at 39 GeV, as well as the partonic mean-field potentials. In our previous studies~\cite{Xu14,Jun16}, these mean-field potentials are taken from a 3-flavor Nambu-Jona-Lasinio (NJL) model, leading to almost zero dynamical mass for partons at high energy densities due to the scalar potential. Since we are only interested in the different spin polarizations of quarks and antiquarks due to vector potentials in the present study, we employ the Lagrangian with only the quark-antiquark vector interaction as well as the external magnetic field for massless partons as follows:
\begin{equation}\label{LL}
  \mathcal{L}=\bar{\psi}\gamma_{\mu}(i\partial^{\mu}-QA_{ext}^{\mu}-\frac{2}{3}G_{V}\langle\bar{\psi}\gamma^{\mu}\psi\rangle)\psi.
\end{equation}
In the above, $\psi=(\psi_{u},\psi_{d},\psi_{s})^{T}$ is respectively the quark field for $u$, $d$, and $s$ quark, $Q={\rm diag}(q_{u}e,q_{d}e,q_{s}e)$ represents respectively their electric charges, and $A_{ext}^{\mu}=(\varphi,\vec{A_{m}})$ is the external electromagnetic potential. The $-\frac{2}{3}G_{V}\langle\bar{\psi}\gamma^{\mu}\psi\rangle$ term represents the flavor-singlet quark-antiquark vector interaction after the mean-field approximation~\cite{Xu14,Jun16}. The vector coupling constant $G_V$, whose value affects the critical point of the chiral phase transition in the phase
diagram~\cite{As89,Fuk08,Car10,Bra13}, is chosen to be 0 or 1.1 times the scalar coupling constant in the original NJL model. The vector density can be expressed as
\begin{equation} \label{vector p}
  \langle\bar{\psi}\gamma^{\mu}\psi\rangle=2N_{c}\sum\limits_{i=u,d,s} \int\frac{d^{3}k}{(2\pi)^{3}E_{i}}k^{\mu}(f_{i}-\bar{f}_{i}),
\end{equation}
where $N_{c}=3$ is the color degeneracy,
$E_{i}=k$ is the energy for massless quarks (antiquarks), and
$f_{i}$ and $\bar{f}_{i}$ are respectively the phase-space
distribution functions of quarks and antiquarks of flavor $i$, which
are calculated from the test-particle method~\cite{Won82} by averaging parallel events in the
dynamical simulation. As in the original NJL model, the above momentum integration is cut off at 750 MeV~\cite{Bra13,Lutz92}.

The Euler-Lagrange equation for each quark flavor $i$ can be obtained from the Lagrangian [Eq.~(\ref{LL})] as
\begin{equation}\label{euler}
  [\gamma^{\mu}(i\partial_{\mu}-A_{\mu})]\psi_{i}=0.
\end{equation}
In the above, $A_{\mu}=(A_{0},-\vec{A})$ contains the time and space
components of the vector potential expressed respectively as
\begin{eqnarray}
   A_{0}&=&b_ig_{V}\rho_{0}+q_ie\varphi, \label{totala0} \\
  \vec{A}&=&b_ig_{V}\vec{\rho}+q_ie\vec{A}_{m}, \label{totala}
\end{eqnarray}
with $g_V=\frac{2}{3}G_V$, $\rho_{0}=\langle\bar{\psi}\gamma^{0}\psi\rangle$ and
$\vec{\rho}\equiv\langle\bar{\psi}\vec{\gamma}\psi\rangle$ being
respectively the time and space components of the vector density, the baryon charge number
$b_i=1$ for quarks and $-1$ for antiquarks, and $q_i$ being the electric
charge number of the quark species $i$. $\varphi$ and $\vec{A}_{m}$ are
the scalar and vector potential of the real external electromagnetic field, and their expressions in vacuum are respectively
\begin{eqnarray}
  \varphi(t,\vec{r})&=&\frac{e}{4\pi}\sum\limits_{n} Z_{n}\frac{1}{R_{n}-\vec{v}_{n}\cdot\vec{R}_{n}}, \label{epotentials}  \\
  \vec{A}_{m}(t,\vec{r})&=&\frac{e}{4\pi}\sum\limits_{n} Z_{n}\frac{\vec{v}_{n}}{R_{n}-\vec{v}_{n}\cdot\vec{R}_{n}}, \label{mpotentials}
\end{eqnarray}
where $Z_{n}$ is the charge number of the $n$th spectator nucleon,
$\vec{v}_{n}$ is its velocity at the retarded
time $t'_{n}=t-|\vec{r}-\vec{r}_{n}|$ when the
radiation is emitted, and $\vec{R}_{n}=\vec{r}-\vec{r}_{n}$ is the relative
position of the field point $\vec{r}$ with respect to the nucleon
position $\vec{r}_{n}$. Considering the finite electrical conductivity of the QGP, the vector potential of the electromagnetic field induced by a point particle with charge $e$ moving in the $+z$ direction at the velocity $v$ along the trajectory $z=vt+z_{0}$
is expressed as~\cite{Krill16}
\begin{eqnarray}\label{eainpla}
 \vec{A}_{m}^e&=&\frac{\hat{z}e}{4\sigma_{con}[(z-z_{0})/v]}\times\frac{\exp\left\{\frac{-b^{2}}{4\{\lambda(t)-\lambda[(z-z_{0})/v]\}}\right\}}{4\{\lambda(t)-\lambda[(z-z_{0})/v]\}} \nonumber  \\
 &\times&\theta[vt-(z-z_{0})]\theta[(z-z_{0})-vt_{0}] \nonumber \\
 &+&\frac{\hat{z}ev\gamma}{4\pi}\int_{0}^{+\infty}dk_{\perp}J_{0}(k_{\perp}b) \nonumber \\
 &\times&\exp[-k_{\perp}^{2}\lambda(t)-k_{\perp}\gamma|(z-z_{0})-vt_{0}|].
\end{eqnarray}
In the above, $t_{0}$ is the time when the QGP emerges, $\sigma_{con}(t)$ is the electrical conductivity of the QGP and $\lambda(t)=\int_{t_{0}}^{t}dt^{'}/[\sigma_{con}(t^{'})]$ is related to its time evolution, $\gamma={1}/{\sqrt{1-v^{2}}}$ is the Lorentz factor, $b$ is the distance between the field point and the point particle with charge $e$ perpendicular to the $z$ direction, $J_0$ is the zeroth-order Bessel function of the first kind,
and $\theta$ is Heaviside step function. Equation~(\ref{mpotentials}) is used to calculate $\vec{A}_{m}$ in vacuum before $t_{0}$, and
Eq.~(\ref{eainpla}) is used to calculate $\vec{A}_{m}$ from the summation of $\vec{A}_{m}^e$ after $t_{0}$ when the QGP is produced. Since partons are continuously produced and $\sigma_{con}$ increases gradually from 0 to finite, $t_0$ should in principle to be set as early as possible. In the present study we choose $t_{0}\sim0.09$ fm/c, before which there are too few partons leading to large fluctuations.

After decoupling the $4\times 4$ Eq.~(\ref{euler}) into the $2\times 2$ Schr\"odinger equation, the single-particle Hamiltonian can be obtained as
\begin{equation}\label{cchamiltonian}
  H=c\vec{\sigma}\cdot\vec{k}+A_{0},
\end{equation}
where $\vec{k}=\vec{p}-\vec{A}$ is the real momentum of the particle with $\vec{p}$ being
its canonical momentum, $c$ is the helicity of the particle,
and $\vec{\sigma}$ are the Pauli matrics. In the semiclassical limit by considering $\vec{\sigma}$ as the expectation value of the particle spin, the canonical equations of motion from the above single-particle Hamiltonian are
\begin{eqnarray}
  \dot{\vec{r}}&=&c\vec{\sigma}, \label{rdot} \\
  \dot{\vec{k}}&=&c\vec{\sigma}\times\vec{B}+\vec{E}, \label{kdot}\\
  \dot{\vec{\sigma}}&=& 2c\vec{k} \times \vec{\sigma},
\end{eqnarray}
where $\vec{B}=\nabla\times\vec{A}$ and $\vec{E}=-\nabla A_{0}-\frac{\partial \vec{A}}{\partial t}$ are the total space and time components of the vector potential, including the contributions from the real electromagnetic field originated from the spectator protons and the effective electromagnetic field originated from the quark-antiquark vector interaction. Using the adiabatic approximation $\vec{\sigma}\approx c\hat{k}-\frac{\hbar}{2k}\hat{k}\times\dot{\hat{k}}$
that satisfies $\hat{k}\cdot \dot{\vec{r}}\approx 1+O(\hbar^{2})$, the chiral kinetic equations of motion can be obtained as~\cite{CKM1,CKM3,CKM4}
\begin{eqnarray}
  \sqrt{G}\dot{\vec{r}}&=\hat{k}+\frac{c\hbar}{2k^{2}}\vec{B}+\frac{c\hbar}{2k^{3}}\vec{E}\times\vec{k}, \label{rdot1} \\
  \sqrt{G}\dot{\vec{k}}&=\hat{k}\times\vec{B}+\frac{c\hbar \vec{k}}{2k^{3}}(\vec{E}\cdot \vec{B})+\vec{E}, \label{kdot1}
\end{eqnarray}
with $\sqrt{G}=1+c\hbar\vec{B}\cdot\vec{k}/(2k^{3})$. Starting from the initial phase-space distribution from the AMPT/HIJING model and with equal numbers of partons having the positive ($c=1$) and negative ($c=-1$) helicity, the partonic phase evolves according to the above chiral kinetic equations of motion. The $\sqrt{G}$ factor in Eq.~(\ref{kdot1}) leads to the modification of the phase-space volume~\cite{gc}, so the statistical value of any observable $X$ is calculated according to $\langle X \rangle = \sum_i X_i \sqrt{G_i} / \sum_i  \sqrt{G_i}$ by taking $\sqrt{G}$ as a weight factor. In the limit of the thermalized
Boltzmann distribution at temperature $T$, the spin polarization of massless spin-$\frac{1}{2}$
fermions is
\begin{eqnarray} \label{theo}
\langle\vec{P}\rangle=\frac{\int \frac{d^{3}\vec{k}}{(2\pi)^{3}}c\dot{\vec{r}}\sqrt{G} \exp(-k/T)}{\int \frac{d^{3}\vec{k}}{(2\pi)^{3}}\sqrt{G} \exp(-k/T)}=\frac{\hbar\vec{B}}{4T^{2}},
\end{eqnarray}
consistent with that from the quantum kinetic approach~\cite{Fang16}. Note here $\vec{B}$ contains the information of the baryon and the electric charge of the particle according to Eq.~(\ref{totala}). On the other hand, $\vec{E}$ in Eqs.~(\ref{rdot1}) and (\ref{kdot1}) is unimportant as it does not contribute explicitly to the spin polarization.

\begin{figure}[ht]
\includegraphics[scale=0.35]{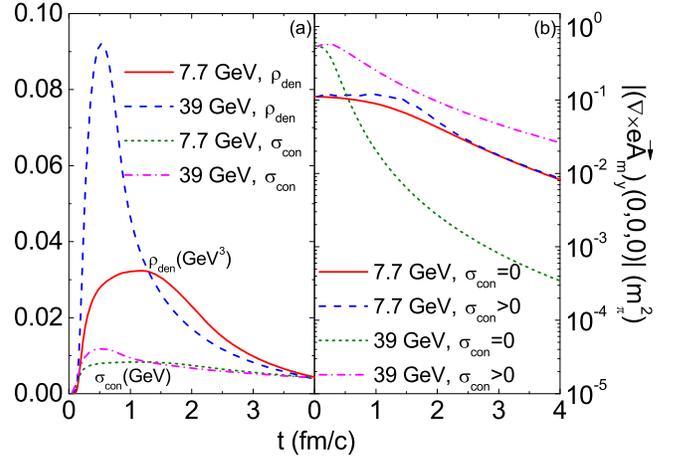}
\caption{(Color online) Left: Time evolution of the total light parton number density
in central cells and the approximated electrical conductivity of the QGP;
Right: Time evolution of the $y$ component of the real magnetic field in vacuum ($\sigma_{con}=0$) and in QGP ($\sigma_{con}>0$) at the center of the collision system. Results are from simulations for midcentral Au + Au collisions.} \label{F1}
\end{figure}

Figure~\ref{F1}(a) displays the time evolution of the total light parton number density $\rho_{den}$
in central cells and the approximated electrical conductivity $\sigma_{con}$ in the QGP formed in
midcentral ($20-50\%$) Au + Au collisions with an average impact parameter 8.87 fm at $\sqrt{s_{NN}}=39$ and $7.7$ GeV from
transport simulations. The temperature dependence of the electrical conductivity is taken as~\cite{Ding11}
\begin{eqnarray}
  \sigma_{con}=0.0058\frac{T}{T_{c}}(\text{GeV}), \label{conductivity}
\end{eqnarray}
where $T$ is the temperature of the QGP and $T_{c}\approx0.165$ GeV is the critical temperature. By assuming that the QGP is a thermalized system consisting massless particles with their momenta following the Boltzmann distribution, the temperature $T$ is extracted from $\rho_{den}$ through the relation $T \approx (\pi^2/24)^{1/3} \rho_{den}^{{1}/{3}}$. Figure~\ref{F1}(b) displays the time evolution of the $y$ component of the real magnetic field in vacuum ($\sigma_{con}=0$) and in QGP ($\sigma_{con}>0$) at the center of the same collision systems. The real magnetic field in vacuum decreases monotonically, and the decreasing trend is sharper at higher collision energies, as found in many similar calculations~\cite{Sk09,Cas11,Deng12}. The strength of the real magnetic field in QGP is slightly enhanced in the early stage compared to that at $t=0$ as a result of the QGP response described by the first term in Eq.~(\ref{eainpla}), and lasts longer compared to that in vacuum, especially at higher collision energies.

\begin{figure}[ht]
\includegraphics[scale=0.35]{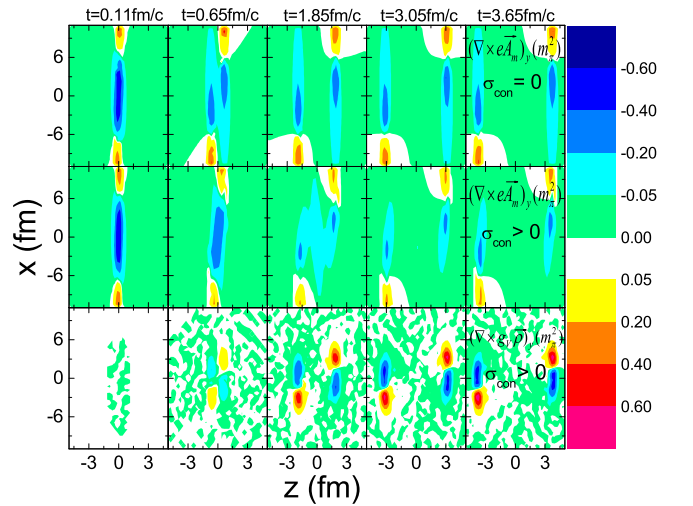}
\caption{(Color online) Contours of the $y$ component of the real magnetic field $(\nabla\times
e\vec{A}_{m})_{y}$ in vacuum (first row), the real magnetic field $(\nabla\times
e\vec{A}_{m})_{y}$ in QGP (second row), and the effective
magnetic field $(\nabla\times g_{V}\vec{\rho})_{y}$ (third row)
in the reaction plane of midcentral Au+Au collisions at $\sqrt{s_{NN}}=39$ GeV. } \label{F2}
\end{figure}

Figure~\ref{F2} displays the spatial distributions of various
fields at different times in the reaction plane ($x$-$o$-$z$ plane) in midcentral Au + Au collisions at $\sqrt{s_{NN}}=39$ GeV. As shown in
the first row, although the real magnetic field in vacuum ($\sigma_{con}=0$) in the central region of the collision system decreases
dramatically with time, the areas where it is strong
move with the spectators towards $\pm z$ directions. The real magnetic field in QGP ($\sigma_{con}>0$) becomes more diffusive especially in the early stage, while the space-time distribution looks similar compared to that in vacuum, as shown in the second row.
Compared with the real magnetic field, the effective magnetic field $(\nabla\times g_{V}\vec{\rho})_{y}$ from the curl of the
net quark flux shows a completely different space-time distribution, which is
positive (negative) at $x\cdot z>0$ ($x\cdot z<0$), as shown in the
third row of Fig.~\ref{F2}. The real and the effective magnetic field are expected to lead to local spin polarizations of various quark species according to their electric and baryon charges.

\begin{figure}[ht]
\includegraphics[scale=0.30]{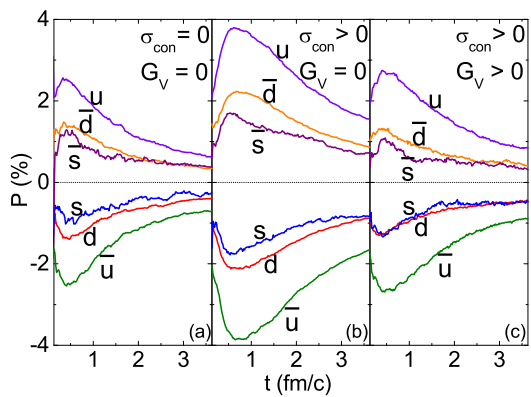}
\caption{(Color online) Time evolution of the spin polarizations of various quark species ($u$, $\bar{u}$, $d$, $\bar{d}$, $s$, and
$\bar{s}$) in the $-y$ direction at midrapidities in midcentral Au + Au collisions at $\sqrt{s_{NN}}=39$ GeV under three different conditions.} \label{F3}
\end{figure}

Figure~\ref{F3} shows the time evolution of the spin polarizations of various quark species under three
different conditions, i.e., with only the real magnetic field in vacuum ($\sigma_{con}=0$, $G_V=0$), with only the real magnetic field in QGP ($\sigma_{con}>0$, $G_V=0$), and with both the real magnetic field in QGP and the effective magnetic field from the quark-antiquark vector interaction ($\sigma_{con}>0$, $G_V>0$). The spin polarizations are in the $-y$ direction parallel to the angular momentum of the collision system. They are calculated according to $P=-(\sum_{n}c_{n}\dot{y}_{n}\sqrt{G_{n}})/(\sum_{n}\sqrt{G_{n}})$,
where the dominating partons with $0.3<\sqrt{G_{n}}<1.7$ are adopted. The common features are as follows.
The spin polarizations are built up quickly in the early stage of the partonic
evolution due to the strong magnetic field in the central zone, then start to decay when the magnetic field in the central zone
becomes weaker, and finally do not change much with time. Even if the magnetic field in the central zone is negligible in the final stage, the spin polarizations are non-zero due to the parton flow induced by the magnetic field.
$u$ quarks have the strongest spin polarization, while the spin polarizations for $d$ and $s$ quarks are much weaker and have an opposite sign compared to that of $u$ quarks, due to their different electric charges. Somehow $s$ quarks have a weaker spin polarization compared with $d$ quarks due to their stiffer momentum distribution in the early stage, while their spin polarizations become similar in the later stage when the system is thermalized due to scatterings. Incorporating the mass of $s$ quarks in the equations of motion by replacing $k$ with $\sqrt{m^2+k^2}$ as in Refs.~\cite{Che14,Sun17} would further reduce the spin polarization of $s$ quarks. For each flavor, quarks and antiquarks have the opposite spin polarization since they have opposite electric and baryon charge numbers. Compared with the spin polarizations in Panel (a) with only the real magnetic field in vacuum, those with only the real magnetic field in QGP shown in Panel (b) are much stronger, as a result of the longer life time of the real magnetic field. As shown in the third row of Fig.~\ref{F2}, the space-time distribution of the effective magnetic field $(\nabla\times g_{V}\vec{\rho})_{y}$ is more profound, adding which leads to the decrease of the spin polarizations for all partons, as shown in Panel (c) of Fig.~\ref{F3}. Note that the space-time distributions of the real and effective magnetic field interact with each other, and their effects on the spin polarizations are not simply additive. The partonic phase ends at about $t \sim 3.7$ fm/c when the partonic scatterings are mostly finished, similar to the parton freeze-out time under the full NJL mean-field potentials~\cite{Guo18}.

\begin{figure}[ht]
\includegraphics[scale=0.30]{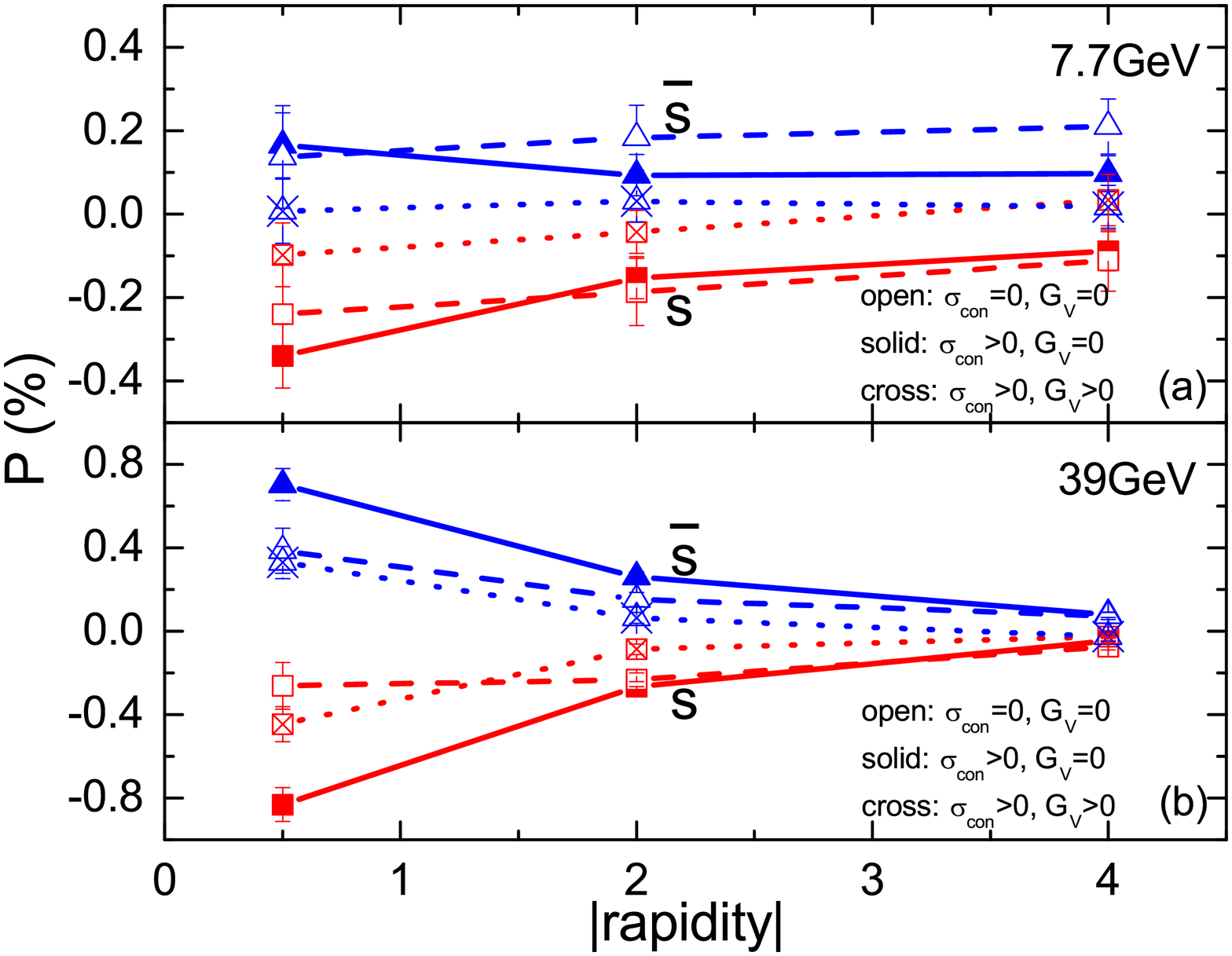}
\caption{(Color online) Rapidity dependence of the spin polarizations of $s$ and
$\bar{s}$ quarks under three different conditions as in Fig.~\ref{F3} at $\sqrt{s_{NN}}=7.7$ (a) and $39$ GeV (b). Note the different $y$-axis scales for 7.7 and 39 GeV.} \label{fig4}
\end{figure}

Figure~\ref{fig4} further displays the rapidity dependence of the spin polarizations of $s$ and $\bar s$
quarks with only the real magnetic field in vacuum ($\sigma_{con}=0$, $G_V=0$), only the real magnetic field in QGP ($\sigma_{con}>0$, $G_V=0$), and both the real magnetic field in QGP and the effective magnetic field from the quark-antiquark vector interaction ($\sigma_{con}>0$, $G_V>0$), in the freeze-out stage of the partonic phase. The spin polarizations are stronger at midrapidities compared to those at larger rapidities, especially at higher collision energies. From the spin-dependent quark coalescence approach, the spin
state of $\Lambda$ ($\bar \Lambda$) is determined by that of its constituent $s$ ($\bar s$)
quark, and the spin polarization value of $\Lambda$ ($\bar \Lambda$) is almost the same as that of the $s$ ($\bar s$) quark~\cite{Sun17}. The larger spin polarization of $\bar s$ than $s$ quarks, leading presumedly to the larger spin polarization of $\bar\Lambda$ than $\Lambda$, is qualitatively consistent with that observed from the STAR data~\cite{Lisa16}. It is seen that the spin polarizations at midrapidities are stronger under the real magnetic field in QGP than those under the real magnetic field in vacuum. However, the quark-antiquark vector interaction reduces the strength of the spin polarization. This means that the uncertainty of $G_V$ may hamper the reliable measure of the magnetic field in relativistic heavy-ion collisions or the electrical conductivity of the QGP through the splitting of the spin polarization between $\Lambda$ and $\bar\Lambda$. Considering the results quantitatively, the splitting of the spin polarization between $\Lambda$ and $\bar\Lambda$ is consistent with that from STAR analysis~\cite{Lisa16} at $\sqrt{s_{NN}}=39$ GeV, while the splitting of the spin polarization at $\sqrt{s_{NN}}=7.7$ GeV is much smaller than that observed experimentally. As shown in Fig.~\ref{F1}, the uncertainty of the magnetic field at $\sqrt{s_{NN}}=7.7$ GeV due to that of the electrical conductivity of the QGP is quite limited. According to Eq.~(\ref{theo}), using even the initial magnetic field and by assuming that the average temperature of the QGP is about $T=0.2$ GeV at $\sqrt{s_{NN}}=7.7$ GeV, the splitting of the spin polarization between $s$ and $\bar s$ quarks is about $0.8\%$, still much smaller compared with that observed experimentally~\cite{Lisa16}.

To summarize, based on the chiral kinetic equations of motion and using the initial parton distribution from a multiphase transport model, we have studied the spin polarizations of various quark species under the real magnetic field induced by the spectator protons as well as the effective magnetic field originated from the quark-antiquark vector interaction. The spin polarizations of partons are found to be much enhanced under the real magnetic field in QGP, compared to the results under the real magnetic field in vacuum, while the effective magnetic field reduces the strength of the spin polarization. A larger spin polarization of $\bar s$ quarks compared to $s$ quarks is observed, leading presumedly to the larger spin polarization of $\bar\Lambda$ than $\Lambda$, while their difference is sensitive to the strength of not only the real magnetic field but also the effective magnetic field. The large splitting of the spin polarization between $\Lambda$ and $\bar\Lambda$ at $\sqrt{s_{NN}}=7.7$ GeV cannot be obtained by the transport simulation or by the thermal estimate at the partonic level.

Our finding challenges the proposal that the splitting of the spin polarization between $\Lambda$ and $\bar\Lambda$ can be a good measure of the magnetic field in relativistic heavy-ion collisions, if the quark-antiquark vector interaction is not well understood. In order to reproduce better the vector meson mass spectrum with the NJL model~\cite{Lutz92}, the strength of the quark-antiquark vector coupling constant $G_V$ is 1.1 times that of the scalar coupling constant $G_s$, as used in the present study. Starting from a color-current interaction and performing the Fierz transformation leads to $G_V=0.5 G_s$ (see, e.g., Ref.~\cite{Vog91}). The elliptic flow splitting between protons and antiprotons as well as that between $K^+$ and $K^-$ at $\sqrt{s}_{NN}=7.7$ GeV favors $0.5G_s<G_V<1.1G_s$~\cite{Xu14}. A recent lattice QCD calculation using
Taylor expansion of the chemical potentials disfavors the critical point at smaller baryon chemical potentials and/or higher temperatures, which seems to favor a large $G_V$~\cite{LQCD}. A large $G_V$ also postpones or even suppresses the appearance of the quark phase in hybrid stars~\cite{Sha13}, and this is favored by the two-solar-mass compact star~\cite{Dem10}. On the other hand, the
spin polarizations of $\Lambda$ and $\bar\Lambda$ may be further modified differently by their spin-orbit couplings in the baryon-rich hadronic phase~\cite{Cse18}, where the real magnetic field is expected to be even weaker. It will be of interest to investigate whether the large splitting of the spin polarization between $\Lambda$ and $\bar\Lambda$ at $\sqrt{s_{NN}}=7.7$ GeV is due to the hadronic evolution through the transport simulation in the future study.

We acknowledge helpful discussions and communications with Che Ming Ko and Xu-Guang
Huang, and thank Chen Zhong for maintaining the high-quality performance of the
computer facility. This work was supported by the Major State Basic
Research Development Program (973 Program) of China under Contract
No. 2015CB856904, the National Natural Science
Foundation of China under Grant Nos. 11475243 and 11421505,
the Shanghai Key Laboratory of Particle Physics and
Cosmology under Grant No. 15DZ2272100.

\end{document}